\documentclass[11pt]{article}
\usepackage{geometry}                
\usepackage{graphicx}
\usepackage{amssymb}
\usepackage{epstopdf}
\usepackage{epsfig}
\newcommand \Pomeron {I\!\!P}
\def\beq{\begin{equation}}
  \def\eeq{\end{equation}}
\title{DGLAP versus perturbative Pomeron  in large  momentum transfer
hard diffractive processes at HERA and LHC.}
\author{
B. Blok\footnote{Email: blok@physics.technion.ac.il} \\
Department of Physics, Technion---Israel Institute of Technology, 32000 Haifa, Israel
\\
L. Frankfurt\footnote{E-mail: frankfur@tauphy.tau.ac.il} \\ School of Physics and Astronomy, \\ Raymond and Beverly
Sackler Faculty of Exact Sciences, \\ Tel Aviv University, 69978 Tel
Aviv, Israel \\  M.Strikman\footnote{E-mail: strikman@phys.psu.edu} \\Physics Department, Penn State University, University Park, PA, USA}

\begin{document}
  \maketitle
\begin{abstract}
We evaluate within the LO DGLAP approximation the dependence on
energy of the cross section of the photo(electro)production of
vector meson V ($V=J/\psi,...$) in the hard elastic processes  off
a parton  $\gamma^*+g\to V=J/\psi+g$  as the function of momentum
transfer $t=(q_{\gamma}-p_V)^2$. We demonstrate that in the limit
$-t \ge m_V^2 +Q^2$ the cross section does not contain double
logarithmic terms in any order of the DGLAP approximation leading
to the energy independent  cross section. Thus the energy
dependence of cross section $\gamma^*+p\to J/\psi+{\rm rapidity
~gap} +X$ is governed at large $t$ by the gluon distribution
within a proton, i.e. it is unambiguously predicted within the
DGLAP approximation including the stronger $W_{\gamma N}$
dependence at larger $-t$ . This prediction explains recent HERA
data. The calculations which follow perturbative Pomeron logic
predict opposite trend of a weaker $W_{\gamma N}$ dependence at
larger $t$. We explain that at the HERA energies double
logarithmic terms characteristic for DGLAP approximation  dominate
in the hard processes as the consequence of constraints due to
energy-momentum conservation. We give predictions for
ultraperipheral hard diffractive processes at the LHC and show
that these processes are well suited for looking for the
contribution of the single logarithmic terms due to the gluon
emission  in  the multiRegge kinematics. We also comment on the
interrelation between energy and $t$ dependence of the cross
sections of the hard exclusive processes.
\end{abstract}

\maketitle \setcounter{page}{1}
\section{Introduction}

\par
The expansion of the pQCD into the domain of the  small $x$ processes  requires
accounting for the  large double logarithmic terms $\alpha_s\log(x_0/x)\log(Q^2/Q^2_0)$.
This can be done within the double logarithmic approximation  (DLA) which follows
from  the DGLAP evolution dynamics.
 Single logarithmic terms $\alpha_s\log(x_0/x)$ arising from gluon radiation in multi Regge
kinematics   are accounted for in the competing perturbative Pomeron approximation.
 The latter  approximation includes the LO double logarithmic
terms also \cite{DGLAP2} but  neglects the running of  the
coupling constant. The interrelation between these approximations
has been understood recently and led to the resummation models
\cite{Ciafaloni, ABF, Salam, lipatovfadin}. In the resummation
models the onset of the perturbative Pomeron occurs at extremely
large energies.

The DGLAP evolution equation gives a good description of the  HERA
data on the structure function $F_{2p}(x,Q^2)$  \cite{Forte} and
the cross sections of hard exclusive processes observed at HERA
\cite{AFS, koepf,BFS}.  In this paper we consider the hard
inelastic diffractive (HID) processes with  large momentum
transfer $-t$ and large rapidity gap, like $\gamma+p\to J/\psi
+{\rm rapidity ~gap} +X$  (Fig.1), that were  studied at HERA
recently \cite{Zeus,H1,galina}. We show that the specific model
independent properties of the  DGLAP approximation which are
absent in the pQCD calculations of Pomeron at large $-t$ ( cf.
refs. \cite{Lipatov, Forshaw}) allow to describe the HERA data.

\par
In the DGLAP approximation the amplitudes are rapidly increasing with energy since
the $log(x_0/x)$ terms that define the energy dependence of the amplitudes, are multiplied
by large logarithms that arise from the integration over parton transverse momenta. Consider
now  the diffraction processes with large
momentum transfer defined above. It was understood  recently
\cite{BFS} that cross section for diffractive vector meson
photoproduction from a parton does not increase with energy  for
$-t\ge M^2_{J/\psi}$ in striking contrast with rapid increase of
this cross section at $t=0$ since logs  arising from the integration
over parton transverse momenta are $\log(M^2_{V}/(Q^2_{0}-t))$, and thus disappear at
$-t\sim M^2_V$. The quantitative theory of such phenomena was
developed in ref. \cite{BFS}, see also brief discussion below.  This
result  is valid in all orders of DGLAP  approximation
and thus  the cross section of diffractive charmonium production off
a parton is energy independent at large $-t$.

\par
\begin{figure}[htbp]
\centerline{\epsfig{figure=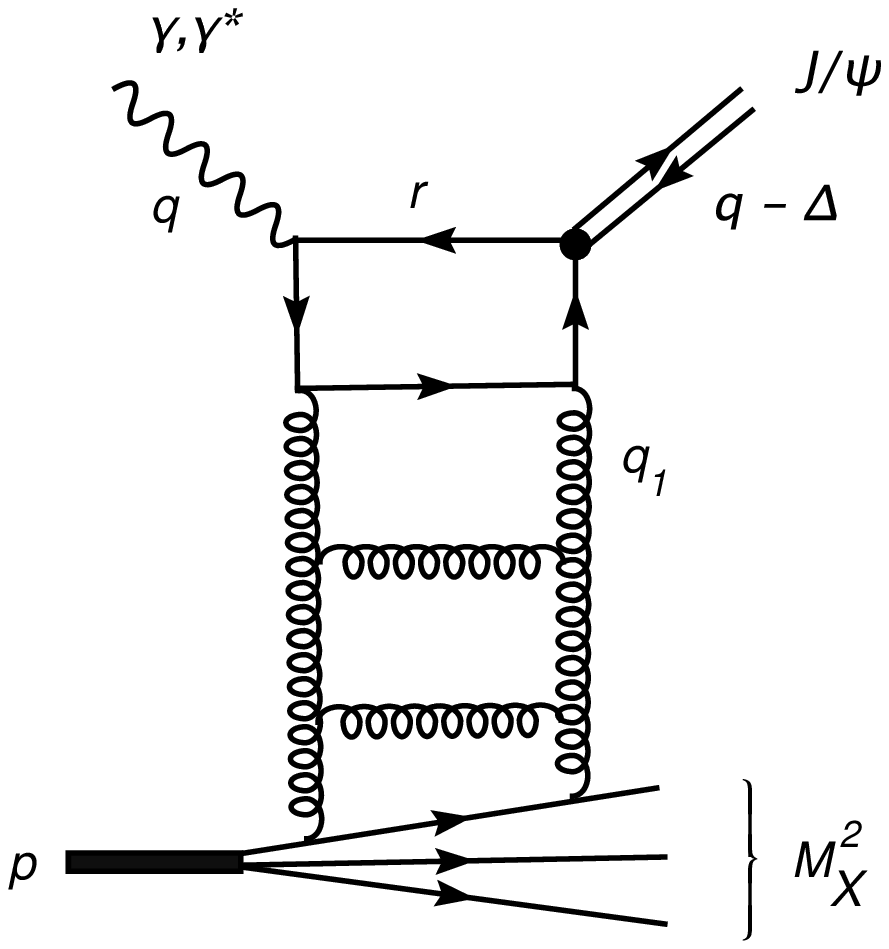,width=15cm,height=15cm,clip=}}
\caption{ The Feynman diagram describing the double diffractive
process in the triple  "Pomeron" limit in pQCD (there is also a cross
diagram, not depicted explicitly.)}\label{S1}
\end{figure}

The  cross section of the HID processes depends on energy at large
$-t$, since the kinematic range of the allowed invariant masses
$M^2_X$ is increasing with energy i.e. due to the  fragmentation
of the  knocked out   parton. Parton fragmentation is described by
the parton  distribution function in a proton,  see Fig. \ref{S1}.
 The dependence of this increase rate on $-t$ is due to the strong dependence
of the rate of increase of the gluon structure function $xG(x,Q^2)$ on $Q^2$. The increase of this rate
with $Q^2$ has been predicted within the DGLAP approximation, and observed
at HERA.  Indeed, if we parametrize
$xG_{p}(x,Q^2)\propto (x_0/x)^{\kappa(Q^2)}$, the exponent  $\kappa (Q^2)$ is increasing with
virtuality $\kappa (Q^2)=
0.048\log(Q^2/\Lambda^2)$ for $x\le 0.01$ , and even more rapidly for larger $x$ \cite{HERA}.
This property is absent within the concept of the Pomeron exchange.
We find that it is this property that  explains the
recent HERA data on large$-t$ diffraction -see Fig.~\ref{fig1}.

\par
The dominance of double logarithmic terms in a wide kinematic
region in the double scale processes, shows that the multiRegge
dynamics could be revealed in the very special one scale processes
where the $Q^2$ evolution is suppressed.  The large $-t$
ultraperipheral processes at LHC represent an example of such
phenomena. We explain that in the kinematical range $-t\ge
M^2_{J/\psi}$ the double logarithmic (DL) terms  are absent. So
the HID phenomena represent a golden  plated process for
uncovering the onset of the pQCD Pomeron. Switching from HERA  to
ultraperipheral processes at LHC significantly increases the
kinematical window allowed for the multiRegge kinematics. In this
case it is possible to select kinematics where a dipole scatters
of partons with rather large $x$, with up to 9 $\div$10 units in
rapidity available for the gluons within the multiRegge
kinematics. An unambiguous signature of these gluons will be
 a rapid increase of the diffractive
cross section with energy in the region where DGLAP predicts the
energy independent cross section.

\par
The amplitude for the  exclusive process: $\gamma+p\to Z+p$  at
nonzero $t$  was studied  in ref. \cite{Bartels} long ago. Recent
derivation of amplitudes of hard diffractive processes within
DGLAP approximation in \cite{BFS} led to the significantly
different dependence of these amplitudes on
$\alpha_s(M^2_{J/\psi}-t)$ and on the collision energy, cf. the
discussion in the end of the paper.

The formulae for the cross sections of diffractive photoproduction of  $J/\psi$  off proton target have
been suggested  also in ref. \cite {Levin} within the dipole model generalized to include screening
corrections and the processes of proton dissociation.  These formulae  differ from the ones which
follow from the QCD  factorization theorem and DGLAP approximation \cite{AFS,BFS}. They do not
include photon scattering  off  a parton and its further fragmentation, and do not correspond to the  triple
"Pomeron" limit. As a result, the obtained dependence on the collision energy and on the hadron mass
produced in  the proton fragmentation significantly differs from the one obtained in \cite{BFS}.
The dependence of QCD evolution on $t$ is in variance with the
dependence of amplitudes on the running coupling constant within the DGLAP approximation
derived in ref. \cite{BFS}. The dependence of cross sections on the incident  energy suggested in the model of
\cite{Levin} contradicts to experimental data, cf. discussion in ref. \cite{galina}.

\par
The letter is organized as follows.  In section II  we compare the predictions of
the DGLAP  approximation for  HID processes with the experimental data at
HERA and show that predicted within DGLAP approximation increase  with $-t$
of the dependence of the cross section on energy is in a good agreement  with the experimental
data. Note,  that the indications of the slowing down of the energy dependence of the cross section
of HID off gluon were first found in the analysis of
experimental data in ref. \cite{FSZ}
where dependence on energy of elastic amplitude for the photo(electro)
production processes off gluon was considered as a  fitting parameter.
In section III we present illustrative estimate of the possible effects of multiRegge
dynamics in the ultraperipheral processes at LHC in the regime  $-t\ge M^2_V$ . In section IV we
discuss the implications of our results for the study of the generalized parton distribution
functions (GPDF).

\section{DGLAP description of HID processes: theory versus experiment.}

\par
The pQCD description of HID processes was recently developed in
ref. \cite{BFS}.
The  differential  cross section in the kinematic range
$-t\le Q^2+M^2_V$ was derived in  ref. \cite{BFS}  and is given by the following formula:
  \footnote[1]{At  moderately large $t$ we perform our
calculation with double logarithmic accuracy only. Hence the scale
associated with the $\gamma \to J/\psi$ vertex is not
unambiguously defined.  The analysis in ref. \cite{koepf} of the
exclusive reaction $\gamma + p \to J/\psi + p$ found  that the
scale is smaller than $m_{J/\psi}^2$ - closer to $Q^2_{eff} \sim 3
\mbox{GeV}^2$. Changing $m_{J/\psi}^2 \to 3 GeV^2$ in the above
formulae
 would result in the applicability of the obtained curves at smaller t.} .
 \beq
\frac{d\sigma}{dt dx_J}=\Phi(t,Q^2,M^2_V)^2\frac{(4N_c^2I_1(u))^2}{\pi
u^2}G(x_J,t).
\label{fe} \eeq
Here
\begin{eqnarray}  u&=&\sqrt{16N_c\log(x/x_J)\chi'}, \, \chi^{'}=\frac{1}{b}\log(\frac{\log((Q^2+M^2_V)/\Lambda^2)}{\log(-t+Q^2_0)/\Lambda^2}),
\nonumber\\[10pt]
x_J&=&-t/(M^2_X-m^2_p-t),x\sim 3(Q^2+M^2_V)/(2s), \, b=11-2/3N_f,N_c=3, s=W^2_{\gamma p}\nonumber\\[10pt]
\label{fe1}
\end{eqnarray}
The factor $\Phi(t,Q^2,M^2_V)$ in eq. \ref{fe}
 is the energy independent function,  which depends on the
details of the wave functions of the  produced quarkonium and a
photon. The second factor corresponds to the distribution of
gluons in a parton, calculated in the DL approximation, The last
factor in eq. \ref{fe} is the gluon structure function of the
nucleon that can be calculated using e.g.  CTEQ6 data (we neglect
small contribution of the quark sea). The function $I_1$ is the
modified Bessel function. In this equation $M^2_X$ is the
invariant mass of the hadronic  system produced due to the
diffractive dissociation of a proton (see Fig.~1),  $\Lambda=300 $
MeV, and $-t$ is the transverse momentum transfer. For the
photoproduction processes we are going to discuss in this letter
the external photon is real: $Q^2=0$ and  $V=J/\Psi$  to ensure
presence of hard scale already for small $t$ and to avoid the end
point contribution which are present for the light vector meson
production. This energy dependence is the same for photoproduction
of longitudinally and transversely polarized onium states.

\par
Let us stress two characteristic features of eq. \ref{fe}. One is
that at small $-t$ close to zero the rate of the  increase of the
cross section with energy  is $\approx$ double of the rate of the
increase of the gluon density.  Second is the absence of the
energy dependence at $-t\ge  M^2_V$  and approximately the same as
for the exclusive $J/\psi$ production.  Indeed, for $t=0$ we
obtain $\chi'=\chi$, i.e. the energy dependence is proportional to
$I_1(u)^2/u^2$, instead $I_1(u)/u$ for the case of the total cross
section, with the same $u$. On the other hand for $-t\ge M^2_V$
and fixed $x_J$  cross section is independent on energy in
arbitrary order of DGLAP approximation: \beq
\frac{d^2\sigma}{dx_Jdt}={\rm const}. \label{pe} \eeq
\par
\begin{figure}[htbp]
\centerline{\epsfig{figure=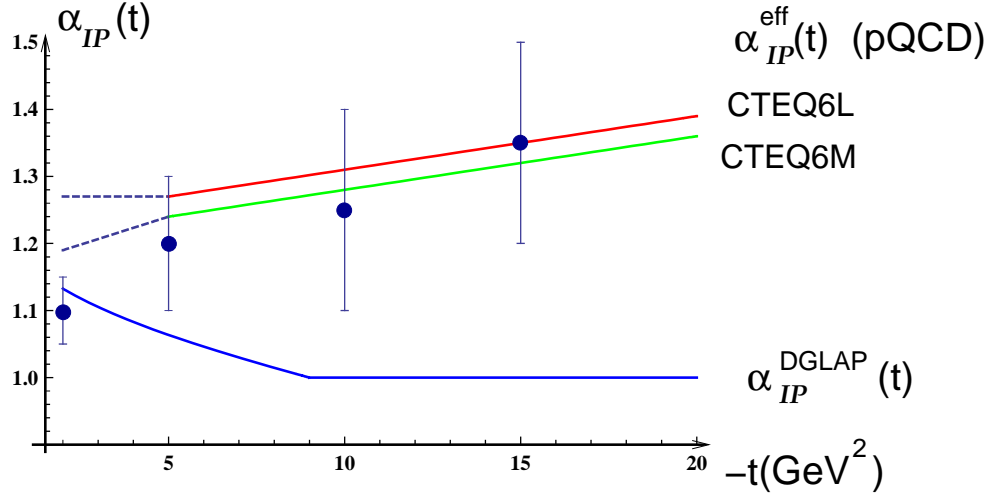,width=13cm,
clip=}} \caption{ The comparison between the experimental data and
theoretical prediction for the HID cross section at HERA for the
"effective Pomeron" $\alpha_P^{\rm eff}(t)$, i.e. (1/2)
logarithmic derivative of the cross section $d\sigma/dt$, obtained
after integrating between the energy dependent cuts, as given in
the text. The dashed curve means large theoretical uncertainties
in the corresponding kinematic region. The values are given at for
$W_{\gamma p}=150$ GeV. In the same figure we depict also "true
(DGLAP) "Pomeron", i.e. logarithmic derivative $\alpha_P(t)^{\rm
DGLAP}$
$\displaystyle{0.5\frac{d(d\sigma/)dtdx_J))}{dlog(x/x_J)}}$ at
this energy. $\Lambda_{\rm QCD}=300$ MeV. }\label{fig1}
\end{figure}

Let us compare  now the theoretical prediction with the recent
experimental data. This comparison, as it was first noted in ref. \cite{FSZ}, is not
straightforward, since  the HERA experiments, see e.g. \cite{galina}, report
the integral over invariant masses:
\begin{eqnarray}
\frac{d\sigma(s,t)}{dt}
&=&\int_{B(s)}^{A(s)} dM^2_X/(M^2_X-t)^2 d^2\sigma/(dtdx_J)(x_J,s,t)\nonumber\\[10pt]
A(s)&=& 0.05s-t, B(s) =1~ \rm{GeV}^2,\nonumber\\[10pt]
 \label{1}
\end{eqnarray}
where we changed the integration variable from $x_J$ to $M^2_X$.
The characteristic feature of eq. \ref{1} is that the dependence
on energy of the cross section comes from two sources. One is the
differential cross section for photon scattering off gluon eq.
\ref{fe}, but also from the dependence on $W_{\gamma N}$ of the
integration limits over $M^2_{X}$.   In particular, when $-t\ge
M^2_{J/\Psi}$ and the energy dependence of the cross section of
photo (electro)production of vector meson from gluon target is
absent, the dependence on energy of the experimentally measured
quantity is given by \beq \frac{d\sigma(s,t)}{dt}
=U(t)\int^{A(s)}_{B(s)}G((-t/(M^2_X-t),t)dM^2_X.\label{12} \eeq
Here $U(t)$ is an energy independent function that can be
expressed through the function $\Phi$ in eq. \ref{fe}. Thus the
kinematic restrictions depend on the collision energy leading to
the energy dependent cross section. As we already mentioned in the
Introduction, both the HERA data and DGLAP results for the gluon
structure  function show, that it increases with the increase of
the gluon virtuality $Q^2$ starting already from $x\sim 0.5$
\cite{HERA}. This property of the structure functions immediately
translates into a steeper energy dependence of $d\sigma/dt$ given
by eq. \ref{12} with the increase of $-t$. In order to compare the
theoretical prediction with the experimental data we calculated
the integral \ref{1} numerically for all $-t$. We use the gluon
structure function given by CTEQ6L and CTEQ6M, leading to only
small differences. We present the main results of our calculations
in Fig.~\ref{fig1}. In order to compare with the experimental data
we calculate the logarithmic derivative \beq
I(s,t)=\frac{1}{2}d\log(d\sigma/dt)/d log(s)\label{21} \eeq for
$s=2\cdot 10^{4}$ GeV$^2$ (we denote this quantity as
$\alpha_{\Pomeron}^{\rm eff}(t)$, and compare it with the data for
$I(s,t)$  presented in  Fig.~9 in ref. \cite{galina}. It is
referred to in ref. \cite{galina} as the "Pomeron" trajectory
$\alpha_\Pomeron(t)-1$.
 We present our results calculated using  eqs. \ref{fe},\ref{12} at Fig.~\ref{fig1} and
compared them with the experimental data as reported in ref.~\cite{galina}.
In the calculation we use CTEQ6M and CTEQ6L gluon structure function $G(x_J,t)$, and
neglect small contribution of  the quark sea.  We depict the curves
for small $-t$ ($-t\sim 2$ GeV$^2$) for such "effective Pomeron" by
the dashed lines,
 since for these $-t$ the integration region
includes the range of $x/x_J\sim 0.1-1$, where nondiagonal (GPD)
effects may be important. In addition, the gluon distribution
function is subject to significant uncertainties.  The results are
clearly within the experimental errors. For comparison we also
depict in Fig.~\ref{fig1} the logarithmic derivative of the double
differential cross section \ref{fe} at this energy, that
corresponds to the "true" Pomeron in the triple Pomeron limit
which enters in eq.~\ref{fe}.

 \par
 Existing calculations of cross sections of large $t$ diffractive processes within
 perturbative Pomeron  hypothesis,  cf \cite{Forshaw} predict qualitatively
 different interplay of $t$ and $W$ dependence. In particular,  in difference
 from the DGLAP approximation  they predicted that dependence on energy of
 gluon structure function of proton should be independent on $Q^2$. Secondly
 perturbative Pomeron "trajectory" weakly depends on $t$ as
   $(\alpha_s(M^2_{V}-t)/\alpha_s(M^2_{V}))\alpha_{\Pomeron}(t=0)$,
  cf. \cite{Lipatov}.   As a result
 perturbative Pomeron calculations predict the decrease with $-t$
 of the energy dependence of the cross section of $\gamma+p\to J/\psi+{\rm rapidity ~gap} +X$.
 This prediction is in variance with the HERA data \cite {galina}.

 \par
Observation that the pQCD Pomeron regime does not set in for  the
HERA kinematics can be understood as the consequence of the
constraints due to the energy-momentum conservation
\cite{FSXX,Schmidt,FRS}. Indeed, the concept of the pQCD Pomeron
is based on the assumption of the dominance of the gluon radiation
in the multiRegge kinematics at high energies.  However there must
be at least 2$\div $2.5 units in rapidity for each gluon
radiation. This means that for the radiation  of even one gluon in
the multiRegge kinematics  a rapidity window of at least 4$\div $5
units in rapidity is required. One should add to this interval the
interval in rapidity  characteristic for the fragmentation
regions. At least 2$\div $2.5 units due to photon fragmentation
plus region in rapidity occupied by DL terms and  larger than
2$\div $2.5 units  interval for the proton fragmentation  (Proton
fragmentation into small masses is suppressed by the power of
$t$). Thus it seems that there is no room in the HERA kinematics
for the gluon radiation in the multiRegge kinematics. On the
contrary  DGLAP approximation conserves longitudinal momentum, and
double logarithmic approximation is a good approximation to DGLAP
formulae at small $x$. The analysis carried in ref.
\cite{Ciafaloni} within the resummation models has found also that
the kernel (the amplitude for scattering of the gluon off a
parton)  is well described within the DGLAP approximation for
reasonable values of $Q^2\sim 20-50$ GeV$^2$, up to $x\sim
10^{-4}\div 10^{-5}$. In fact such a good agreement between theory
and experiment can be considered as an experimental verification
of the energy-momentum constraints on the gluon radiation within
the multiRegge kinematics.

\section{MultiRegge gluons in HID processes with large momentum transfer $-t$ at LHC.}

\begin{figure}[htbp]
\centerline{\epsfig{figure=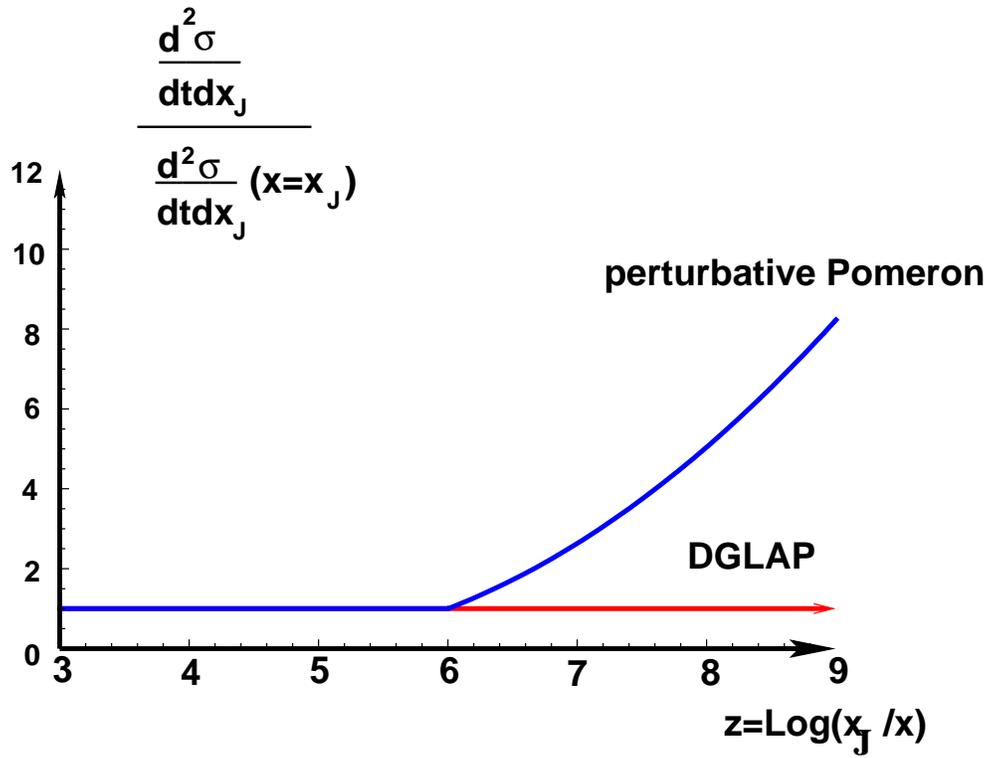,width=13cm,
clip=}} \caption{ The increase of the cross section
$d\sigma/)dtdx_J$ with energy ($z=Log(x/x_J)$) at LHC in DGLAP and
perturbative "Pomeron" scenarios (for fixed $x_J$)}\label{fig2}
\end{figure}

Consider now the ultraperipheral processes at the LHC (for a review see \cite{Baltz:2007kq}). In
this case one  may have up to $14$ units in rapidity,  i.e. up to 9 units
in rapidity  may be available for a ladder describing gluon-parton scattering.
As we have discussed above, for moderate
$-t$ a significant part of this kinematic window will be filled by
gluon radiation leading to  the double logarithms.
Different steps of the diffractive process occupy different regions of rapidity:
first interval is fragmentation of virtual photon or heavy quarkonium which
is dominated by gluon radiation within double logarithmic kinematics.
This range corresponds to the effective change  of
transverse momenta from $\sim (M^2_V-t)/4$ in the impact factor to
$\sim -t/4$. The rest of the evolution with energy will be
determined by the radiation of gluons within the multi Regge
kinematics , corresponding to a single-scale
process, whose amplitude  does not contain double logarithms.A simple estimate suggests that for small $-t$ most of the
kinematic range (even at the LHC) will be dominated  by double logarithmes. For larger
$-t$ the precise kinematic range dominated  by double logarithms would be smaller and can
be estimated. However, we shall not need it, if we are
interested in model independent
signature of gluons radiated in the multi Regge kinematics.
Indeed, as it is clear from the previous subsection for $-t\ge M^2_V$ the double
logarithmic terms are absent, and the entire
increase of the double differential cross section $d^2\sigma/(dtdx_J)$
will be due to multi Regge gluons. In Fig.~\ref{fig2} we showed
(for illustrative purposes only, using currently popular pQCD
Pomeron models with intercept $\beta (t)-1\sim 0.25$,
and proportional to $\alpha_s(M^2_{J/\Psi}-t)$, the behavior of the
$d^2\sigma/(dtdx_J)$ as a function of energy. We shall see a rapid
increase of cross section starting from $log(x_J/x)\ge 5$. The
absence of such a rise will be a sign that multi Regge dynamics
does not appear up to an onset of a black limit, when the pQCD is
not applicable any more.
\par
Note, that comparing a rate of a rise of cross section in the
ultraperipheral processes on nucleon and nucleus, it is possible to find
the onset of the black disk regime \cite{FSZ1}.

\section{Implications for   the  studies of GPDs in
small x exclusive processes}

The necessity to account for the DL terms in the QCD evolution at small x
has important implications for the interpretation of the energy dependence of the $-t$ slope
for exclusive vector meson production, DVCS  studied at HERA.

Indeed the factorization theorem \cite{Collins:1996fb} which
allows to express the amplitude through the convolution of the
photon and the vector meson  wave functions and the generalized
parton distribution (GPD) was proven in the limit $(Q^2 +M^2_V)\gg
-t$. In the limit when $-t$ is compared to $(Q^2+M^2_V)$ it
requires modifications because of disappearance of the difference
in the scales.

 For $-t $  close to zero  the DGLAP evolution of $\alpha'$  which is
 present at the boundary condition
 leads to a rather  slow variation of $\alpha'$ with $Q^2$ \cite{Frankfurt:2003td}.

To be able to compare with experimental data for $\alpha'_{eff}$
we approximate the amplitude as $A(t=0, s_0) \exp(\alpha'_{eff}
t\ln (s/s_0))$ and estimate:
\begin{equation}
\alpha'_{eff} \sim (\alpha(0)-1) /Q^2.
\end{equation}
Note that the analysis  of the $J/\psi $ exclusive photoproduction
suggests the effective scale of the order  $Q_{eff}^2 \sim 3~
\mbox{GeV}^2$ leading to
\begin{equation}
\alpha'_{eff} \sim 0.07 ~\mbox{GeV}^{-2}. \label{est}
\end{equation}

 The estimate of eq.\ref{est}  should be compared to the experimental numbers measured
 at HERA by ZEUS\cite{Chekanov:2004mw} and H1\cite{Aktas:2005xu} collaborations which
 agree well with each other.  For example, in case of photoproduction ZEUS  reported
 $\alpha' \sim 0.115 \pm 0.018 ~\mbox{GeV}^{-2}$ with $\alpha_{\Pomeron}$ for  the last two
 points at $-t = 1.0, 1.3 ~\mbox{GeV}^2$ consistent with one.  It is of interest also that the
 ZEUS and H1 experimental data do not contradict a  a  smaller value of
$\alpha'  $ for electroproduction of $J/\psi$. In particular ZEUS
reports for $Q^2= 6.8 ~\mbox{GeV}^2$ - $\alpha' \sim 0.07 \pm
0.05~ \mbox{GeV}^{-2}$ which is consistent with the pattern
expected for the discussed effect.

The studies of 3D image of the nucleon at small x require in the exclusive $J/\psi$ production
require reaching $-t \sim 2 \mbox{GeV}^2$ in order to probe the gluon density at small impact
parameters. It follows from our discussion that reaching the range of applicability of the
factorization theorem \cite{Collins:1996fb} will require $Q^2$ of the order of 10 GeV$^2$.

\section{Conclusion}

We have shown that DGLAP predictions are in a good agreement with
the behavior of HID processes observed at HERA. We found that the
ultraperipheral collisions at LHC  are a unique place where the
onset of  gluon radiation in the  multiRegge kinematics may be
observed in the near future.

\newpage
\clearpage
\clearpage

\end{document}